\documentclass[sigconf]{acmart}
\usepackage{amsmath}
\usepackage{amssymb}
\usepackage{mathrsfs}
\usepackage{algorithm}
\usepackage{color,xcolor}
\usepackage{algpseudocode}
\usepackage{amsmath}
\usepackage{graphics}
\usepackage{epsfig}
\usepackage{balance}
\usepackage{threeparttable}
\usepackage{enumitem}
\usepackage{bm}
\usepackage{verbatim}
\usepackage{soul}
\usepackage{multirow}
\usepackage{makecell}
\usepackage{xspace}
\usepackage{appendix}

\newcommand{\eg}{\emph{e.g.,}\xspace}
\newcommand{\ie}{\emph{i.e.,}\xspace}

\AtBeginDocument{%
  \providecommand\BibTeX{{%
    \normalfont B\kern-0.5em{\scshape i\kern-0.25em b}\kern-0.8em\TeX}}}

\makeatletter
\g@addto@macro\normalsize{%
\abovedisplayskip 2pt plus 1pt minus 1pt%
\belowdisplayskip \abovedisplayskip
\abovedisplayshortskip 2pt plus 1pt minus 1pt%
\belowdisplayshortskip 2pt plus 1pt minus 1pt%
}
\makeatother

\copyrightyear{2020}
\acmYear{2020}
\setcopyright{acmcopyright}\acmConference[CIKM '20]{Proceedings of the 29th ACM International Conference on Information and Knowledge Management}{October 19--23, 2020}{Virtual Event, Ireland}
\acmBooktitle{Proceedings of the 29th ACM International Conference on Information and Knowledge Management (CIKM '20), October 19--23, 2020, Virtual Event, Ireland}
\acmPrice{15.00}
\acmDOI{10.1145/3340531.3412729}
\acmISBN{978-1-4503-6859-9/20/10}
\begin{document}

\fancyhead{}

%%
%% The "title" command has an optional parameter,
%% allowing the author to define a "short title" to be used in page headers.

\title{MTBRN: Multiplex Target-Behavior Relation Enhanced Network for Click-Through Rate Prediction}

% yf before:
% \title{MTBRN: Bridging and Associating User Behaviors with Target Item for Click-Through Rate Prediction}

%%
%% The "author" command and its associated commands are used to define
%% the authors and their affiliations.
%% Of note is the shared affiliation of the first two authors, and the
%% "authornote" and "authornotemark" commands
%% used to denote shared contribution to the research.
\author{Yufei Feng$^{1\ast}$, Fuyu Lv$^{1\ast}$, Binbin Hu$^2$, Fei Sun$^1$, Kun Kuang$^3$, Yang Liu$^4$, Qingwen Liu$^1$, Wenwu Ou$^1$}\thanks{$\ast$ denotes equal contributions.}

\affiliation{%
\institution{$^1$Alibaba Group, Hangzhou, China} 
\institution{$^2$Ant Financial Services Group, Hangzhou , China}
\institution{$^3$Zhejiang University, Hangzhou, China}
\institution{$^4$Indiana University Bloomington, Indiana, United States}
}
\email{
fyf649435349@gmail.com,
fuyu.lfy@alibaba-inc.com,
bin.hbb@antfin.com,
ofey.sf@alibaba-inc.com,
kkun2010@gmail.com,
yl82@indiana.edu,
{xiangsheng.lqw, santong.oww}@alibaba-inc.com
}

%%
%% By default, the full list of authors will be used in the page
%% headers. Often, this list is too long, and will overlap
%% other information printed in the page headers. This command allows
%% the author to define a more concise list
%% of authors' names for this purpose.
% \renewcommand{\shortauthors}{Trovato and Tobin, et al.}

%%
%% The abstract is a short summary of the work to be presented in the
%% article.
\begin{abstract}
    Click-through rate (CTR) prediction is a critical task for many industrial systems, such as display advertising and recommender systems. Recently, modeling user behavior sequences attracts much attention and shows great improvements in the CTR field. Existing works mainly exploit attention mechanism based on embedding product when considering relations between user behaviors and target item. However, this methodology lacks of concrete semantics and overlooks the underlying reasons driving a user to click on a target item. In this paper, we propose a new framework named \textbf{M}ultiplex \textbf{T}arget-\textbf{B}ehavior \textbf{R}elation enhanced \textbf{N}etwork (MTBRN) to leverage multiplex relations between user behaviors and target item to enhance CTR prediction. Multiplex relations consist of meaningful semantics, which can bring a better understanding on users' interests from different perspectives. To explore and model multiplex relations, we propose to incorporate various graphs (\eg knowledge graph and item-item similarity graph) to construct multiple relational paths between user behaviors and target item. Then Bi-LSTM is applied to encode each path in the path extractor layer. A path fusion network and a path activation network are devised to adaptively aggregate and finally learn the representation of all paths for CTR prediction. Extensive offline and online experiments clearly verify the effectiveness of our framework.
\end{abstract}


%%
%% The code below is generated by the tool at http://dl.acm.org/ccs.cfm.
%% Please copy and paste the code instead of the example below.
% \begin{CCSXML}
% <ccs2012>
% <concept>
% <concept_id>10002951.10003317.10003347.10003350</concept_id>
% <concept_desc>Information systems~Recommender systems</concept_desc>
% <concept_significance>500</concept_significance>
% </concept>
% <concept>
% <concept_id>10010147.10010257.10010293.10010294</concept_id>
% <concept_desc>Computing methodologies~Neural networks</concept_desc>
% <concept_significance>500</concept_significance>
% </concept>
% </ccs2012>
% \end{CCSXML}

% \ccsdesc[500]{Information systems~Recommender systems}
% \ccsdesc[500]{Computing methodologies~Neural networks}

%%
%% Keywords. The author(s) should pick words that accurately describe
%% the work being presented. Separate the keywords with commas.
% \keywords{datasets, neural networks, gaze detection, text tagging}

%%
%% This command processes the author and affiliation and title
%% information and builds the first part of the formatted document.
\maketitle

\section{Introduction}
    \begin{figure}
        \centering
        \includegraphics[scale=0.31]{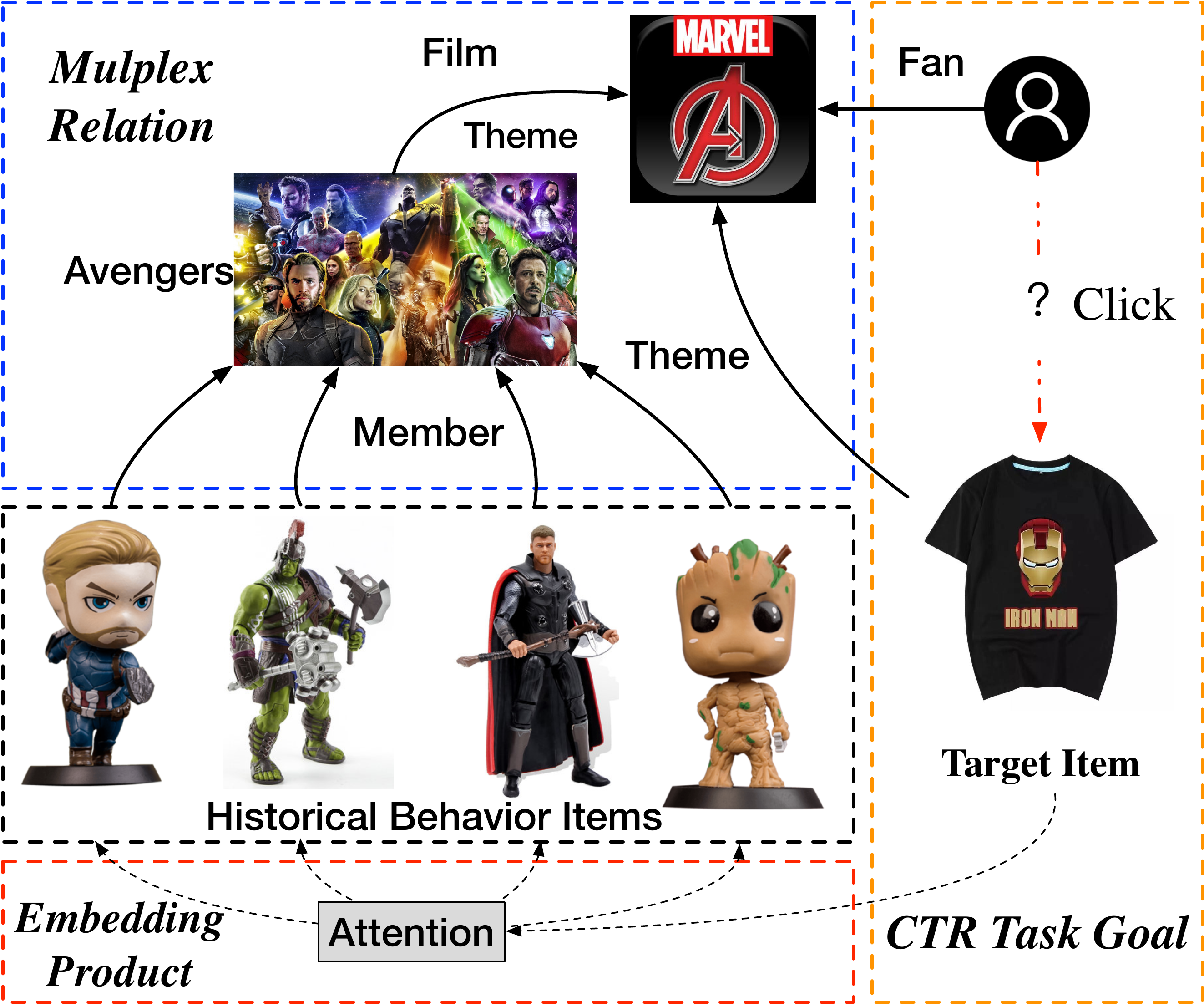}
        \caption{An illustration example of multiplex relations between a user's historical behaviors and the target item in CTR prediction task.}
        \label{fig:illustration}
    \end{figure}

    Click-through rate (CTR) prediction lives the heart at display advertising and recommender systems. It estimates the probability of a user to click on a given target item. The quality of CTR is fundamental to user experience and user retention. Recently, modeling user behavior sequences is prevailing in CTR prediction. Several related algorithms~\cite{Zhou:DIN,Feng:DSIN,Pi:MIMN,zhou2019deep,ni2018perceive,ren2019lifelong} have been proposed and achieved good performance in real-world applications. They represent behavior sequences as fixed-length vectors of user interests and feed them into the deep neural network for final CTR prediction. These models mainly exploit attention mechanism based on embedding product to aggregate user behaviors w.r.t target item. 
    
    Although these methods have achieved performance improvement to some extent, they still face a few major weaknesses. Most importantly, they lack of concrete semantics and overlook the underlying reasons driving a user to click on a target item. As a result, they fail to precisely comprehend a user’s interest. As illustrated in Figure~\ref{fig:illustration}, a movie fan of \textit{Avengers} is likely to click on an "Iron Man" graphic T-shirt, even if he has only browsed some spin-off products of Avengers. Such underlying reason, also named multiplex relation, is composed of multi-typed semantic relatedness (\eg "fan of", "member of" and "theme of") and different types of items or entities (\eg "members of Avengers" and "film of \textit{Avengers}"). It is difficult for existing models to capture the multiplex relation where these spin-off products and T-shirts can be explicitly linked to the same theme of \textit{Avengers} by meaningful semantics. Furthermore, there are more than one multiplex relations between user behaviors and target item. These multiplex relations are particularly helpful to reveal the preferences (\ie reasons) of users on consuming items from different perspectives. For instance, after watching the movie \textit{Avengers: Infinity War}, a user may choose either \textit{Avengers: End Game} or \textit{Sherlock Holmes} to watch next. Because the former is the sequel of \textit{Avengers: Infinity War}, and the latter is also starred by \textit{Robert Downey Jr.} The two movies should be recommended to the user since both relations with \textit{Avengers: Infinity War} must be considered. Therefore, without explicitly modeling multiplex relations, it is conceptually difficult for precise recommendation.
    
    In order to address aforementioned problems, we propose a new framework named \textbf{M}ultiplex \textbf{T}arget-\textbf{B}ehavior \textbf{R}elation enhanced \textbf{N}etwork (MTBRN) for CTR prediction. MTBRN transfers the information of multiplex relations between user behaviors and target item in a unified framework, so as to bridge and associate them. Knowledge graph (KG) emerges as an alternative to describe such relations, as the surge of interests in incorporating KG into recommender systems due to its comprehensive auxiliary data~\cite{zhang2016collaborative,Wang:KPRN,Wang:KGCN,Wang:KGAT,Hu:McRec,wang2018ripplenet,huang2018improving,wang2019multi,wang2018dkn,yu2014personalized,wang2019explainable,cao2019unifying,Wang2019KGN}. It introduces semantic relatedness among items and various entities, which can capture multiple underlying connections between items. That motivates us to model multiplex relations based on KG. 
    
    We explore and construct multiple paths between user behaviors and target item on KG to capture multiplex relations via graph search algorithm. In a modern web-scale recommender system, there are other graphs that can provide useful linking information to describe multiplex relations. An item-item similarity graph, for instance, can establish high-order connection between similar items. In the remaining part of this paper, we demonstrate our framework can incorporate these graphs in the same way as KG. To integrate those relational paths into MTBRN, we use Bi-LSTM to encode each path. Relational paths from various graphs can benefit and complement each other, so we devise a fusion network for their higher order feature interaction. After the fusion network, different path representations are adaptively aggregated into the final representation of multiplex relations through an attention based activation network. At last, the representation and other features are concatenated and fed into feature interacting layer for CTR prediction. Experiments were conducted on a proprietary industrial dataset and a public dataset, on which our framework displays state-of-the-art results. MTBRN has been fully deployed into product recommender of one popular E-commerce Mobile App and achieves significant CTR improvements by 7.9\%.
    
    The main contributions of this paper are summarized as follows:
    \begin{itemize}
        \item We highlight the importance of multiplex relations between user behaviors and target item in CTR prediction. We propose a path based method to leverage such relations on different graphs. 
        \item A new CTR prediction framework named MTBRN is proposed to explore and model multiplex relations. Multiple relational paths are extracted from various graphs. Bi-LSTM and a path fusion/activation network are employed to adaptively learn the final representation of multiplex relations.
        \item We performed extensive experiments on a proprietary industrial dataset and a public dataset. Experimental results verify the rationality of each graph and the effectiveness of the proposed MTBRN framework.
    \end{itemize}

\section{Problem Formulation} \label{section:pf}

\begin{figure*}
    \begin{center}
    \includegraphics[scale=0.5]{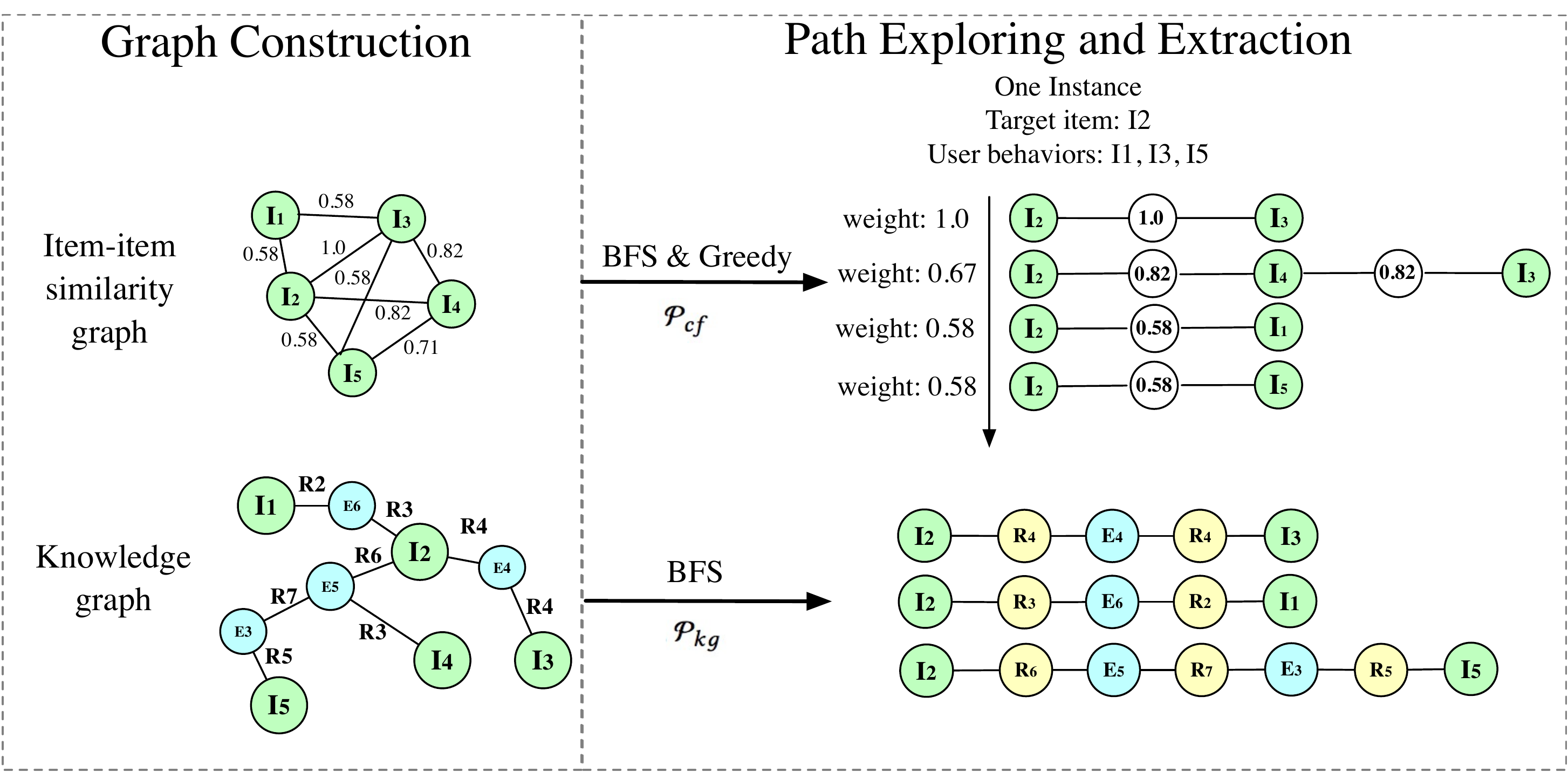}
    \end{center}
    \caption{The graph construction methods and path exploration and extraction strategy.}
    \label{fig:gcpe}
\end{figure*}

In this section, we formulate the problem of click-through rate prediction with multiplex relations between user behaviors and target item. Specifically, we construct the following two graphs to extract multiplex relations, which are illustrated in the left part of Figure~\ref{fig:gcpe}.

\textbf{Item-item similarity graph.} 
The item-item similarity graph is denoted as $\mathcal{G}_{cf} = \{(i, sim(i, j), j)|i, j \in \mathcal{I}\}$, where $\mathcal{I}$ is the set of items and $sim(i, j)$ describes the similarity between item $i$ and $j$. We calculate $sim(i, j)$ via the idea of item-based collaborative filtering~\cite{sarwar2001item}, which can be formulated as follows,
\begin{equation}
    \begin{split}
    sim(i, j) = \frac{\mathbf{Y}[:,i]\odot \mathbf{Y}[:,j]}{\sqrt{\left\| \mathbf{Y}[:,i] \right\|_2  \left\| \mathbf{Y}[:,j]  \right\|_2} }.
    \end{split}
\end{equation}
Here, $\odot$ refers to element-wise product, $\textbf{Y} \in \mathbb{R}^{m \times n}$ is the user-item interaction matrix from users' historical behaviors, where $m$ and $n$ is the number of users and items respectively and $y_{ui} = 1$ indicates that user $u$ interacted with item $i$, $0$ otherwise. For example, the triple $(i, 0.3, j)$ indicates that the similarity of item $i$ and item $j$ is 0.3. The triple $(j, 0.9, k)$ indicates that the similarity between item $j$ and item $k$ is $0.9$. The item $j$ is the junction of the two triples.

\textbf{Knowledge graph.}
The knowledge graph describes semantic correlation among items and real-world entities via various relations, which can be denoted as $\mathcal{G}_{kg} = \{(h, r, t)|h,t \in \mathcal{E}, r \in \mathcal{R}\}$, where $\mathcal{E}$ is the set of items and entities, and $\mathcal{R}$ is the set of relations. For example, the triple $(i, category, longuette)$ indicates that the item $i$ belongs to the $longuette$ category. The triple $(longuette, parent, clothing)$ indicates the parent of $longuette$ is clothing. The $longuette$ entity is the junction of the two triples. 

\begin{table}
  \caption{Notations.}
  \label{table:notations}
  \begin{tabular}{ll}
    \toprule
    Notations & Description\\
    \midrule
    $y_{uv}, \hat{y}_{uv}$ & \makecell[{}{p{5.8cm}}]{label and the predicted probability}\\
    $u$, $v$ & \makecell[{}{p{5.8cm}}]{the user and target item}\\
    $\mathcal{U}$, $\mathcal{I}$ & \makecell[{}{p{5.8cm}}]{the user set and item set}\\
    $x_u$, $x_v$, $\mathcal{B}$ & \makecell[{}{p{5.8cm}}]{user profile, item profile and user behaviors}\\
    $b_i$ & \makecell[{}{p{5.8cm}}]{the i$_{\mathit{th}}$ user behavior item}\\
    $\mathcal{G}_{\mathit{cf}}$, $\mathcal{G}_{\mathit{kg}}$ & \makecell[{}{p{5.8cm}}]{item-item similarity graph and knowledge graph}\\
    $\mathcal{P}_{\mathit{kg}}$, $\mathcal{P}_{\mathit{cf}}$ & \makecell[{}{p{5.8cm}}]{path sets extracted from $\mathcal{G}_{\mathit{cf}}$ and $\mathcal{G}_{\mathit{kg}}$}\\
    $\textbf{v}_i$ & \makecell[{}{p{5.8cm}}]{embedding vector for the i$_{\mathit{th}}$ feature}\\
    $\textbf{e}_i$ & \makecell[{}{p{5.8cm}}]{embedding vector for the i$_{\mathit{th}}$ node of the path}\\
    $\textbf{h}_{p}$ & \makecell[{}{p{5.8cm}}]{the concatenated output of Bi-LSTM}\\
    $\textbf{H}_{\mathit{cf}}$, $\textbf{H}_{\mathit{kg}}$ & \makecell[{}{p{5.8cm}}]{the output of $\mathcal{P}_{\mathit{cf}}$, $\mathcal{P}_{\mathit{kg}}$ in the relational path extractor layer}\\
    $\textbf{H}_{\mathit{fu}}$ & \makecell[{}{p{5.8cm}}]{the output of $\textbf{H}_{\mathit{cf}}$, $\textbf{H}_{\mathit{kg}}$ in the relational path fusion layer}\\
    $\textbf{A}_{\mathit{cf}}$, $\textbf{A}_{\mathit{kg}}$, $\textbf{A}_{\mathit{fu}}$& \makecell[{}{p{5.8cm}}]{the output of $\textbf{H}_{\mathit{cf}}$, $\textbf{H}_{\mathit{kg}}$ and $\textbf{H}_{\mathit{fu}}$ in the relational path activation layer}\\
    \bottomrule
  \end{tabular}
\end{table}

\textbf{Multiplex relations enhanced CTR prediction.}
Now, we formulate the CTR prediction problem to be addressed in this paper. We assume a set of historical click records between users and items, denoted as $\mathcal{Y}$. $\mathcal{Y}$ is comprised of $\{u, v, \mathcal{B}, y_{uv} | u \in \mathcal{U}, v \in \mathcal{I}, y_{uv} \in \{0, 1\}\}$, where  $\mathcal{U}$ and $\mathcal{I}$ respectively represent the user and item sets, $\mathcal{B}$ is the set of user behaviors consisting of item ids that the user has recently clicked on and $y_{uv}$ is set to one if and only if user $u$ has clicked on item $v$. Moreover, each user $u$ is associated with a user profile $x_u$ consisting of sparse features (\eg user id and gender) and numerical features (\eg user age), while each target item $v$ is also associated with a item profile $x_v$ consisting of sparse features (\eg item id and brand) and numerical features (\eg price).
%where $x_u$ denotes the user profile consisting of sparse features (\eg user id and gender) and numerical features (\eg user age), $x_v$ represents target item profile consisting of sparse features(\eg item id and brand) and numerical features (\eg price), $x_b$ is the set of user behaviors consisting of item ids that the user has recently clicked. $y \in \{0, 1\}$ is set to one if and only if user $u$ has clicked item $i$. 
In order to effectively explore and exploit the multiplex relations between user behaviors and target item, we elaborately construct knowledge graph $\mathcal{G}_{\mathit{kg}}$ and item-item similarity graph $\mathcal{G}_{\mathit{cf}}$ to enhance CTR prediction. Formally, our goal is to learn a prediction function $\hat{y}_{uv} = \mathcal{F}(x_u, x_v, \mathcal{B}, \mathcal{G}_{kg}, \mathcal{G}_{cf}; \Theta)$, such that $\hat{y}_{uv}$ represents the predicted probability of user $u$ to click on target item $v$ and $\Theta$ represents the parameters of the prediction function $\mathcal{F}$. The notations are summarized in Table~\ref{table:notations}.

\section{MTBRN Framework}
In this section, we introduce our proposed framework MTBRN. We first propose the path-based algorithm that captures and models multiplex relations from different graphs. Afterwards, we elaborate on the deep neural network architecture of MTBRN, which is proposed to encode the relational information of the paths extracted from auxiliary graphs and adaptively learn how paths contribute to the final prediction.

\subsection{Path Exploration and Extraction Strategy}\label{PEESSec}
In this part, we introduce the strategy to effectively explore and extract paths between user behaviors and target item, which is a natural way to describe multiplex relations on graphs. Previous path-based models either use the random walk strategy~\cite{Wang:KPRN} or design an auxiliary task (\eg matrix factorization) to assign weights to paths~\cite{Hu:McRec}. 
Unfortunately, random walk based methods may harm the stability of model performance while auxiliary task based methods only work under specific settings.
To effectively capture multiplex relations between user behavior $b_i \in \mathcal{B}$ and target item $v$, we adopt different search strategies to extract paths on different graphs, listed as follows:
\begin{itemize}
    \item For the item-item similarity graph $\mathcal{G}_{\mathit{cf}}$, we exhaustively
    search any potential path following breadth-first search (BFS) and greedy-selection principle according to the similarity score. We only keep top $k_{\mathbf{G}_{cf}}$ paths with the shortest length.
    One demo path extracted from the item-item similarity graph is defined as: $b_i \rightarrow sim(b_i, A) \rightarrow A \rightarrow \cdots \rightarrow sim(\cdot, v) \rightarrow v$, where ($b_i$, $sim(b_i, A)$, $A$) is one triple in $\mathcal{G}_{\mathit{cf}}$.
    \item For the knowledge graph $\mathcal{G}_{\mathit{kg}}$, we follow the BFS principle
    to generate all paths over the linked
    relations and entities between user behaviors and target
    item, and keep top $k_{\mathcal{G}_{kg}}$ paths with the shortest length. One demo path extracted from the knowledge graph can be defined as: $b_i \rightarrow r_1 \rightarrow e_1 \rightarrow \cdots \rightarrow v$, where ($b_i$, $r_1$, $e_1$) is one triple in $\mathcal{G}_{\mathit{kg}}$.
\end{itemize}

We illustrate the procedure in the Figure~\ref{fig:gcpe}. Afterwards, we can get two types of path sets: $\mathcal{P}_{\mathit{cf}}$ and $\mathcal{P}_{\mathit{kg}}$, which capture multiplex relations between user behaviors and target item from different perspectives.

\subsection{Model Architecture}

\begin{figure*}
    \begin{center}
    \includegraphics[scale=0.33]{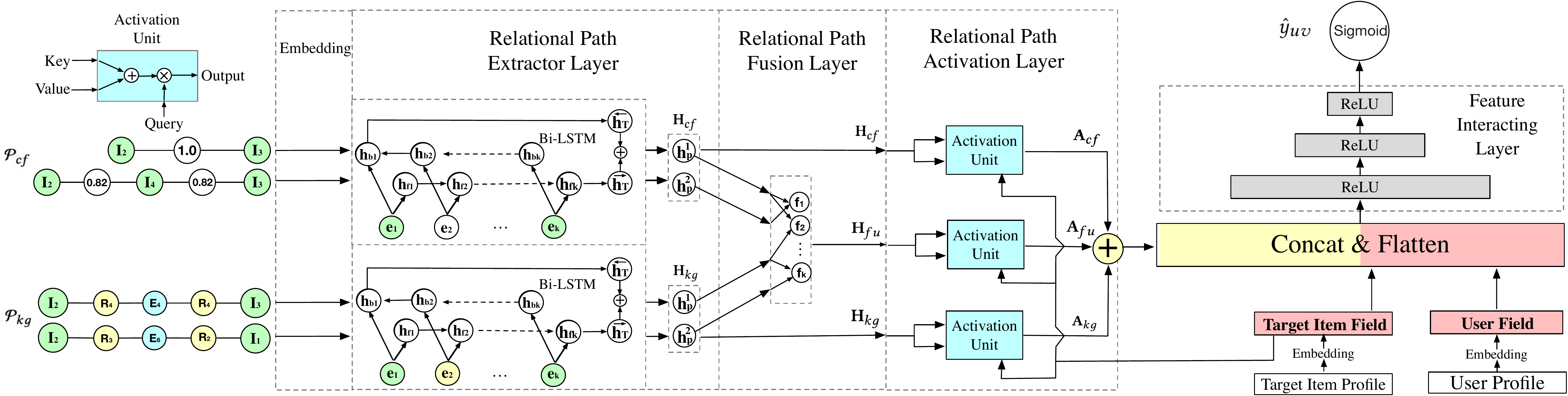}
    \end{center}
    \caption{Overview of proposed framework MTBRN. The right part is embedding vectors of user and target item profile features. The left part is our main contribution, which processes the extracted relational paths. We use Bi-LSTM, path fusion and activation network to encode multiplex relational paths. Representations from the two parts are concatenated, flattened and then fed into feature interacting layer for the final prediction.}
    % Overall, before feature interacting layer, MTBRN is composed of two parts. First, we use Bi-LSTM in the relational path extractor layer to encode relational paths from different graphs separately. Afterwards, the higher-order interactions of different paths are fully explored in the relational path fusion layer. Finally, the attention mechanism w.r.t. the target item is applied in the relational path activation layer. 
    \label{fig:MTBRN}
\end{figure*}

As shown in Figure~\ref{fig:MTBRN}, MTBRN consists of two parts before feature interacting layer. The right part is the embedding vectors transformed from the user profile feature and target item profile feature.
The left part models the extracted relational paths, which is composed of three layers from left to right: (1) \textit{relational path extractor layer} extracts bi-directional relational information on paths; (2) \textit{relational path fusion layer} captures the higher-order interactions of paths; (3) \textit{relational path activation layer} adaptively learns the representation of the relational paths w.r.t. the target item. Finally the outputs of the two parts are concatenated, flattened and fed into the feature interacting layer for the final prediction. Next, we will present detailed illustration of the proposed MTBRN model.

\textbf{Embedding.} 
Embedding is a popular technique that projects each feature to a dense vector representation. Formally, let $\textbf{v}_i \in \mathbb{R}^{d}$ be the embedding vector for the i$_{\mathit{th}}$ feature. Since there exist numerical features (\eg price) in original features, we rescale the embedding vector by its input feature value to account for the real valued features. 
In this way, the embedding of the i$_{\mathit{th}}$ feature for the input vector $\mathbf{x}$ is calculated as $\mathbf{v}_i\mathbf{x}_i$. Due to the sparsity of the input feature, we only need to preserve the embeddings for non-zero features. Thus, the final embedding of input feature vector $\mathbf{e}$ is obtain as:
\begin{equation} 
    \begin{aligned}
    \mathbf{e} =\mathbf{v}_1\mathbf{x}_1 \oplus \mathbf{v}_2\mathbf{x}_2 \oplus \dots \oplus \mathbf{v}_i\mathbf{x}_i\dots, \ \ \ \ \mathbf{x}_i \neq 0,
    \end{aligned}
\end{equation}
where $\oplus$ is the concatenation operation. Following the above embedding procedure, the user and target item profile feature space can be represented by $\textbf{x}_u$ and $\textbf{x}_v$, respectively. One path in $\mathcal{P}_{\mathit{cf}}$ and $\mathcal{P}_{\mathit{kg}}$ can be uniformly represented by $[b_i; node_1; \dots; node_j; \dots; v]$, where $b_i$ is the $i$-th user behavior item, $v$ is the target item, and $node_j$ can be either sparse features (\ie items, relations and entities), or numerical features (\ie similarity score). With embedding, the path can be generally represented by [$\textbf{e}_1$; \dots ; $\textbf{e}_k$], where $k$ is the length of the path.
% {\itshape User Behavior} representation $\textbf{x}_b$ is obtained by average pooling over embeddings of user behavior items. Specifically

% \subsection{Relational Path Extractor Layer} user-target
\textbf{Relational Path Extractor Layer.} The path sets describe the multiplex relations of target item and user behaviors from different perspectives, and this layer is designed to extract the information passing along the paths. Previous path-based models either use CNN~\cite{Hu:McRec} or LSTM~\cite{Wang:KPRN} to encode the relational paths from user to target item. Whereas items in the user behavior and target item are in the same semantic space and the transmission of information along the path is always asymmetric. Therefore, we naturally apply Bi-LSTM~\cite{Graves:bilstm} to extract two-way information transmitted on the path asymmetrically. Mathematically, LSTM~\cite{Hochreiter:LSTM} network is implemented as follows:
\begin{equation} 
    \begin{aligned}
    \textbf{i}_t &= \sigma(\textbf{W}_{xi}\textbf{e}_t + \textbf{W}_{hi}\textbf{h}_{t-1} + \textbf{W}_{ci}\textbf{c}_{t-1} + \textbf{b}_i)\\
    \textbf{f}_t &= \sigma(\textbf{W}_{xf}\textbf{e}_t + \textbf{W}_{hf}\textbf{h}_{t-1} + \textbf{W}_{cf}\textbf{c}_{t-1} + \textbf{b}_f)\\
    \textbf{c}_t &= \textbf{f}_{t}\textbf{c}_{t-1} + \textbf{i}_{t}\tanh(\textbf{W}_{xc}\textbf{e}_t + \textbf{W}_{hc}\textbf{h}_{t-1} + \textbf{b}_{c})\\
    \textbf{o}_t &= \sigma(\textbf{W}_{xo}\textbf{e}_t + \textbf{W}_{ho}\textbf{h}_{t-1} + \textbf{W}_{co}\textbf{c}_{t} + \textbf{b}_o)\\
    \textbf{h}_t &= \textbf{o}_{t}\tanh(\textbf{c}_t)\\
    \end{aligned}
    \label{eq:bi-lstm}
\end{equation}
where $\sigma(\cdot)$ is the logistic function, $\textbf{i}$, $\textbf{f}$, $\textbf{o}$ and $\textbf{c}$ are the input gate, forget gate, output gate and cell vectors, respectively. Forward and backward LSTMs model the bi-direction information, that is, the last representation of each path $\textbf{h}_{p}$ is calculated as follows:
\begin{equation}
    \begin{split}
    \textbf{h}_{p} = \overrightarrow{\textbf{h}_{T}} \oplus \overleftarrow{\textbf{h}_{T}}
    \end{split}
\end{equation}
where $\overrightarrow{\textbf{h}_{T}}$ and $\overleftarrow{\textbf{h}_{T}}$ represent the last hidden state of the forward LSTM and backward LSTM, respectively. Note that the parameters of Bi-LSTM are shared when encoding paths from the same set. After the relational path extractor layer, $\mathcal{P}_{\mathit{cf}}$ and $\mathcal{P}_{\mathit{kg}}$ are represented by $\textbf{H}_{\mathit{cf}}$ and $\textbf{H}_{\mathit{kg}}$, respectively. For example, $\textbf{H}_{kg} = [\textbf{h}^{1}_{p}; \dots; \textbf{h}^{i}_{p}; \dots]$ denotes last representations of all paths in $\mathcal{P}_{\mathit{kg}}$.

\textbf{Relational Path Fusion Layer.} Relational paths can benefit and complement each other. The previous two paths $b_1 - 0.6 - v$ and $b_1 - category - longuette - category - v$, for example, may bring much more information to light if considered together. Inspired by the improvements~\citeN{Rendle:FM,Juan:FFM,Cheng:WDL,Guo:DeepFM} of feature interaction in the CTR field, we further capture the higher order interaction among representations of paths. Mathematically, the interactive representation $\textbf{H}_{\mathit{fu}}$ can be calculated as follows:
\begin{equation}
    \textbf{H}_{fu} = \{\textbf{h}^{i}_{p}\cdot\textbf{h}^{j}_{p} \mid 1 \leq i \leq k, i + 1 \leq j \leq k \} 
\end{equation}
where $\textbf{h}^{i}_{p} \in \textbf{H}_{cf} \cup \textbf{H}_{\mathit{kg}}$, $\cdot$ is element-wise multiply and $k = k_{\mathcal{G}_{cf}} + k_{\mathcal{G}_{kg}}$ is the number of all paths.

\textbf{Relational Path Activation Layer.} Intuitively, relational paths contribute unequally to target item and further influence the final prediction. Taking two paths $b_1 - 0.6 - v$ and $b_2 - 0.3 - v$ in $\mathcal{P}_{\mathit{cf}}$ as an example, $v$ is more similar to target item $b_1$ than $b_2$ since the first path has the higher similarity score than the second path. Meanwhile, relational paths extracted from different graphs are not on the same scale. Taking another two paths $b_1 - 0.6 - v$ and $b_1 - category - longuette - category - v$ for example, it is hard to distinguish which one is more effective. Therefore, the weights of path representations in $\textbf{H}_{cf}, \textbf{H}_{kg}, \textbf{H}_{\mathit{fu}}$ need to be reassigned w.r.t. the target item. For this reason, the attention mechanism~\cite{Dzmitry:Attention} is applied to conduct alignment between paths and the target item. Mathematically, the adaptive representation of relational paths in each path set w.r.t. the target item is calculated as follows:
\begin{equation}
    \begin{split}
    a^i &= \frac{\exp(\textbf{h}^{i}_{p}\textbf{W}\textbf{x}_v))}{\sum_{j=1}^{n} \exp(\textbf{h}^{j}_{p}\textbf{W}\textbf{x}_v)} \\
    \textbf{A} &= \sum_{i=1}^{n} a^i\textbf{h}^{i}_{p}\\
    \end{split}
\end{equation}
where $n$ is the number of paths in each path set and \textbf{W} is the trainable parameters. After the relational path activation layer, $\textbf{H}_{\mathit{cf}}$, $\textbf{H}_{\mathit{kg}}$ and $\textbf{H}_{\mathit{fu}}$ respectively are encoded into vectors $\textbf{A}_{\mathit{cf}}$, $\textbf{A}_{\mathit{kg}}$ and $\textbf{A}_{\mathit{fu}}$.

% \subsection{Feature Interacting Layer}
\textbf{Feature Interacting Layer.} Following previous studies~\citeN{Zhou:DIN,Feng:DSIN,Pi:MIMN,zhou2019deep} in the CTR prediction field, Multiple Layer Perceptron (MLP) is applied for better feature interaction. Here we calculate the final output as follows:
\begin{equation}
    \begin{split}
    \hat{y}_{uv} = \sigma(f(f(f(\textbf{x}_u \oplus \textbf{x}_v \oplus \textbf{A}_{cf} \oplus \textbf{A}_{kg} \oplus \textbf{A}_{fu}))))
    \end{split}
\end{equation}
where $f(\textbf{x}) = ReLU(\textbf{W}\textbf{x} + \textbf{b})$, $\sigma(\cdot)$ is the logistic function and $\hat{y}_{uv}$ represents the prediction probability of the user $u$ to click on the target item $v$.

% \subsection{Loss Function}
\textbf{Loss Function.} We reduce the CTR prediction task to a binary classification problem with binary cross-entropy loss function, which can be defined as follows:
\begin{equation} 
    \begin{split}
    Loss = - \frac{1}{N} \sum_{(u, v) \in \mathbb{D}} (y_{uv}\,\log \hat{y}_{uv} + (1 - y_{uv})\,\log(1 - \hat{y}_{uv}))
    \end{split}
    \label{eq:loss function}
\end{equation}
where $\mathbb{D}$ is the training dataset and $y_{uv} \in \{0, 1\}$ represents whether the user $u$ clicked on the target item $v$.

\section{Experiments}
We evaluate the proposed framework MTBRN on a proprietary industrial E-commerce dataset and the public Yelp dataset. Moreover, we conduct strict online A/B testing to evaluate the performance of MTBRN after deployed to real-world settings. Specifically, we will make comprehensive analyses about MTBRN, with the aim of answering the following questions:
\begin{itemize}
    \item \textbf{RQ1} How does MTBRN perform compared with other state-of-the-art (SOTA) user behavior enhanced CTR models? 
    \item \textbf{RQ2} How does MTBRN perform compared with competitors that can leverage the same graph in our framework for recommendation?
    \item \textbf{RQ3} How do the multiplex relations between user behaviors and target item derived from different graphs benefit CTR prediction?
    % \item \textbf{RQ3} How can MTBRN provide reasonable explanations for the final prediction?
\end{itemize}

\subsection{Datasets and Graph Description}
We report detailed description of the two datasets and all graphs utilized by MTBRN in Table~\ref{table:t2}.

\subsubsection{E-commerce Dataset.} It is an industrial real-world recommender dataset collected from a popular E-commerce Mobile App. The dataset consists of impression/click logs in 8 consecutive days, where clicked ones are treated as positive instances and negative otherwise. Logs from 2019-08-22 to 2019-08-28 are used for training, and logs from 2019-08-29 are for testing. Moreover, E-commerce dataset contains user profile (\eg id, age, and gender), item profile (\eg id, category, and price) and real-time user behaviors\footnote{Real-time user behaviors refer to user behaviors before this instance occurs.}.

\subsubsection{Yelp Dataset.} Yelp datasets records interactions between users and local business and contains user profile (\eg id, review count, and fans), item profile (\eg id, city, and stars) and real-time user behaviors\footnote{https://www.yelp.com/dataset.}. To adapt for the CTR prediction task, we treat all observed interactions as positive instance. For each user-item pair in positive instance, we randomly sample 5 negative samples that have no interaction record with the specific user to constitute negative instance set. Then, in chronological order, we take each user's last 30 instances for testing and last 31 to 120 instances for training. 

\subsubsection{Item-item Similarity Graph.} We construct the item-item similarity graph as introduced in section~\ref{section:pf}. To model the real-world recommender system and prevent information leakage, we only construct the graph based on user behaviors that do not exists in the training and testing datasets. Moreover, considering the tremendous number of items, we only keep top 5 neighbors with highest similarity score for each item and the max depth of each node is set to 3.

\subsubsection{Knowledge Graph.} 
Knowledge-aware recommendation relies highly on the quality of the knowledge graph. We construct the knowledge graph following procedures described in~\citeN{luo2020alicoco}. For the E-commerce dataset, relations include category, parent, season, style, etc. For the Yelp dataset, relations include category, location, attribute, etc.

\begin{table}%[H]
  \caption{Statistics of the dataset and graphs.}
  \label{table:t2}
  \begin{tabular}{ccc}
    \toprule
    Description & E-commerce & Yelp\\
    \midrule
    Users & 0.2 billion & 45.4 thousand\\
    Items & 0.1 billion & 45.1 thousand\\
    Records & 7.2 billion & 1.0 million\\
    User behaviors & 10 & 10\\
    \hline
    Triplets in $\mathcal{G}_{\mathit{cf}}$ & 26.2 billion & 8.1 million\\
    \hline
    Relations in $\mathcal{G}_{\mathit{kg}}$ & 34 & 35 \\
    Entities in $\mathcal{G}_{\mathit{kg}}$ & 14.4 million & 83.3 thousand \\
    Triplets in $\mathcal{G}_{\mathit{kg}}$ & 33.8 billion & 1.6 million \\
    \bottomrule
  \end{tabular}
\end{table}

\subsection{Experimental Setup}
\subsubsection{Competitors}
We consider two kinds of representative CTR prediction  methods: user behavior sequence enhanced methods (\ie YoutubeNet, DIN, DIEN and DSIN), item-item similarity graph based method (\ie GIN) and knowledge graph based methods (\ie RippleNet, KPRN and KGAT). To examine the effect of the multiplex relations derived from different graphs and relational path fusion layer, we prepare three variants of MTBRN (\ie MTBRN$_{cf}$, MTBRN$_{kg}$ and MTBRN$_{r}$). The competitors are given below:

\begin{itemize}
    \item \textbf{YoutubeNet}~\cite{Paul:YoutubeNet} is designed for video recommendation in Youtube. It treats user behaviors equally and applies average pooling operation.
    \item \textbf{DeepFM}~\cite{Guo:DeepFM} is technically designed to capture the multi-order interactions of features. It combines FM and deep model, without the need of complicated feature engineering.
    \item \textbf{DIN}~\cite{Zhou:DIN} uses the embedding product attention mechanism to learn the adaptive representation of user behaviors w.r.t. the target item.
    \item \textbf{DIEN}~\cite{zhou2019deep} designs an auxiliary network to capture user's temporal interests and proposes AUGRU to model the interest evolution.
    \item \textbf{DSIN}~\cite{Feng:DSIN} divides the user behavior sequence into multiple sessions and designs the extractor layer and evolving layer to extract the session interests and model how they evolves during time.
    \item \textbf{GIN}~\cite{Li:GIN} is the first to mine and aggregate the user's latent intention on the co-occurrence item graph with graph attention technique. GIN can be easily applied on item-item similarity graph. 
    \item \textbf{RippleNet}~\cite{wang2018ripplenet} explores the multiple \textit{ripples} of user behaviors on the knowledge graph and propagates the representation of the target item recursively layer by layer.
    \item \textbf{KPRN}~\cite{Wang:KPRN} applies LSTM to directly model the multiple user-item paths via the knowledge graph and then aggregate them for the final prediction.
    \item \textbf{KGAT}~\cite{Wang:KGAT} recursively propagates the high-order connectivity of the user and item via the knowledge graph and user-item bipartite graph with graph attention technique. 
    \item \textbf{MTBRN}$_{\mathit{cf}}$ : MTBRN with only item-item similarity graph.
    \item \textbf{MTBRN}$_{\mathit{kg}}$ : MTBRN with only knowledge graph.
    \item \textbf{MTBRN}$_r$ : MTBRN without relational path fusion layer.
\end{itemize}

\subsubsection{Evaluation Metrics} In our experiments, we evaluate the performance of different methods for comparison via AUC (Area Under ROC Curve) and Logloss (cross entropy), which are widely adopted in the CTR field. The larger AUC, the better performance. 
Base on our practice lessons, 0.1\% increase of offline AUC on our proprietary dataset is corresponding to relative 1\% online CTR lift. 

\subsubsection{Implementation.}
We implemented all the models in Tensorflow 1.4. We tailored models which were not originally designed for the CTR prediction task, including concatenation with other features and addition of MLP layers at last. We did not apply pre-training, batch normalization and regularization techniques. Instead, random uniform initializer is employed. 
With the computational cost in mind, only 2 layers of neighbours are reserved for each user behavior item for GIN. For KGAT, the neighbour depth of the user and item is set to 5 and 4, respectively. For RippleNet, the depth of ripple is set to 3. The extracted paths on the item-item similarity graph and knowledge graph are reserved up to 50.
All models are tuned using Adagrad optimizer with learning rate 0.001 and batch size 300. Embedding size of each feature is set to 4. The hidden units in MLP layers are set 512, 256, and 128, respectively.
We ran each model three times and computed the mean to eradicate any discrepancies.

\subsection{Performance Comparison (RQ1\&RQ2)}\label{sec:exp1}

\begin{table}%[H]
  \caption{Model performance (AUC and Logloss) with each separate graph and all graphs.}
  \label{table:model_performance}
  \begin{tabular}{c|c|c|c|c|c}
    \hline
    \multirowcell{2}{Graph} & \multirowcell{2}{Model} & \multicolumn{2}{{c|}}{E-commerce} & \multicolumn{2}{{c}}{Yelp} \\ 
    \cline{3-6}
    {} & {} & {AUC} & {Logloss} & {AUC} & {Logloss} \\
    \hline
    - & YoutubeNet & 0.6017 & 0.6279 & 0.7109 & 0.4899 \\
    - & DeepFM & 0.6037 & 0.6192 & 0.7334 & 0.4882\\
    - & DIN & 0.6058 & 0.5735 & 0.7520 & 0.4579 \\
    - & DIEN & 0.6065 & 0.5643 & 0.7581 & 0.4518\\
    - & DSIN & 0.6073 & 0.5394 & 0.7774 & 0.4392\\
    \hline
    \multirow{3}{*}{$\mathcal{G}_{\mathit{cf}}$} & GIN & 0.6073 & 0.5416 & 0.7604 & 0.4471\\
    & MTBRN$_{cf}^\dag$ & 0.6094 & 0.5329 & 0.7915 & 0.4129\\
    & MTBRN$_{cf}^\ddag$ & \textbf{0.6103} & \textbf{0.5244}& \textbf{0.7936} & \textbf{0.4075}\\
    \hline
    \multirow{4}{*}{$\mathcal{G}_{\mathit{kg}}$} & RippleNet & 0.5975 & 0.6369 & 0.7324 & 0.4844\\
    & KGAT & 0.6062 & 0.5624 & 0.7876 & 0.4214 \\
    & KPRN & 0.6091 & 0.5292 & 0.8267 & 0.3897 \\
    & MTBRN$_{\mathit{kg}}$ & \textbf{0.6209} & \textbf{0.4628} & \textbf{0.9088} & \textbf{0.3486}\\
    \hline
    \multirow{2}{*}{ALL} & MTBRN$_{r}$ & 0.6235 & 0.4509 & 0.9231 & 0.3213\\
    & MTBRN & \textbf{0.6246} & \textbf{0.4482} & \textbf{0.9408} & \textbf{0.3058}\\
    \hline
  \end{tabular}
\begin{tablenotes}
\item[$\dag$ ($\ddag$)] $\dag$ ($\ddag$) Paths shorter than 5 (7) are reserved, exploring up to 2 (3) layers of neighborhood.
\end{tablenotes}
\end{table}

In this section, we start off comparing the performance of MTBRN with SOTA user behavior enhanced CTR models, as well as with other competitors leveraging each auxiliary graph. We report the performance of all models on the two datasets\footnote{Note that the relative improvements on the public Yelp dataset are much higher than those on the industrial E-commerce dataset, because the negative samples of the public Yelp dataset are generated by random sampling, which means easier to distinguish.} in Table~\ref{table:model_performance}.

\subsubsection{User behavior Enhanced CTR models.} As shown in Table~\ref{table:model_performance}, DIN improves AUC obviously by leveraging the attention mechanism to activate the user's relevant interests w.r.t. the target item. DIEN achieves better performance with the technically designed auxiliary net and AUGRU to model the interest evolution. DSIN performs better than DIEN by extracting user's session interests. Nevertheless, most competitors in the CTR field reallocate weights to user behaviors only based on the item embedding with the attention mechanism. It can hardly figure out the complex reasons driving the user to click the target item. Overall, MTBRN significantly outperforms above state-of-the-art competitors on both datasets, which mainly benefits from two aspects: (1) the extracted multiplex relational paths from different graphs are more reasonable and concrete, so as to provide powerful clues why the user will click on the target item; (2) the technical design of MTBRN helps capture the multiplex relations of user behaviors and the target item. Both contribute much to the final prediction and help achieve the best performance. To answer \textbf{RQ2}, we gave extensive insights on how each graph and different components of MTBRN contribute to the best performance. More complicated models are used as competitors.

\subsubsection{Item-item Similarity Graph.}
As shown in Table~\ref{table:model_performance}, GIN outperforms DIN, mainly benefiting from the exploration of users' latent intention in graphs. However, GIN still ignores the relation between user behaviors and target item as well as the linked similarity score between items.
In MTBRN$_{cf}^\dag$, we explore and model the paths between user behaviors and the target item within the range of 2-layer adjacent neighborhood as the same as GIN. MTBRN$_{cf}^\dag$ outperforms GIN in both datasets. It empirically demonstrates the usefulness of the relation paths between user behaviors and target items for CTR prediction.
Furthermore, we flexibly extend the neighbour depth of the graph to 3 (\ie MTBRN$_{cf}^\ddag$) for exploiting high-order information on the graph and, not surprisingly, more improvement is observed on MTBRN$_{cf}^\ddag$. It conveys the message that longer paths can capture higher-order similarities of items and benefit the final prediction in the long run.

\subsubsection{Knowledge Graph.} 
We present the AUC performance of various knowledge-aware models for the CTR prediction task on both datasets in Table~\ref{table:model_performance}. KGAT and KPRN both outperform DIN, and specifically KPRN offers more increase on the both datasets. In contrast, RippleNet renders inferior performance than DIN. One reasonable explanation is that modeling relations explicitly between user behaviors and target item is more efficacious than user preference propagation in the KG.
KGAT incorporates knowledge graph and user-item bipartite graph, hence it yeilds more improvements compared with DIN. However, KGAT has been found empirically to involve useless information for the final prediction and fails to capture direct interaction between user behaviors and target item. KPRN benefits from reasonable and explainable user-target paths derived from knowledge graph and outperforms KGAT. MTBRN$_{\mathit{kg}}$ devotes to capturing the reasonable and explainable knowledge relations of user behaviors and target item. Moreover, the relational path extractor layer and activation layer help obtain the representation of multiplex relational paths and activate those related to the target item. Therefore, MTBRN$_{\mathit{kg}}$ outperforms other competitors with the same knowledge graph.

\subsubsection{Effect of Relation Path Fusion Layer.} 
We conduct extensive experiments to verify the effectiveness of the proposed relational path fusion layer.
We report the detailed comparison of model performance of MTBRN with or without paths fusion (\ie MTBRN$_r$ and MTBRN) in Table~\ref{table:model_performance}. 
Not surprisingly, MTBRN$_r$ outperforms MTBRN with single graph data (\ie MTBRN$_{\mathit{cf}}$ and MTBRN$_{\mathit{kg}}$) since it incorporates multiplex relations derived from different graphs for the final prediction. 
Moreover, MTBRN performs better than MTBRN$_r$, which demonstrates the effectiveness of the proposed relational path fusion layer. 

\subsection{Validity Analysis of Paths (RQ3)}\label{sec:RQ1}

\begin{figure}%[htbp]
    \begin{flushleft}
    \includegraphics[scale=0.127]{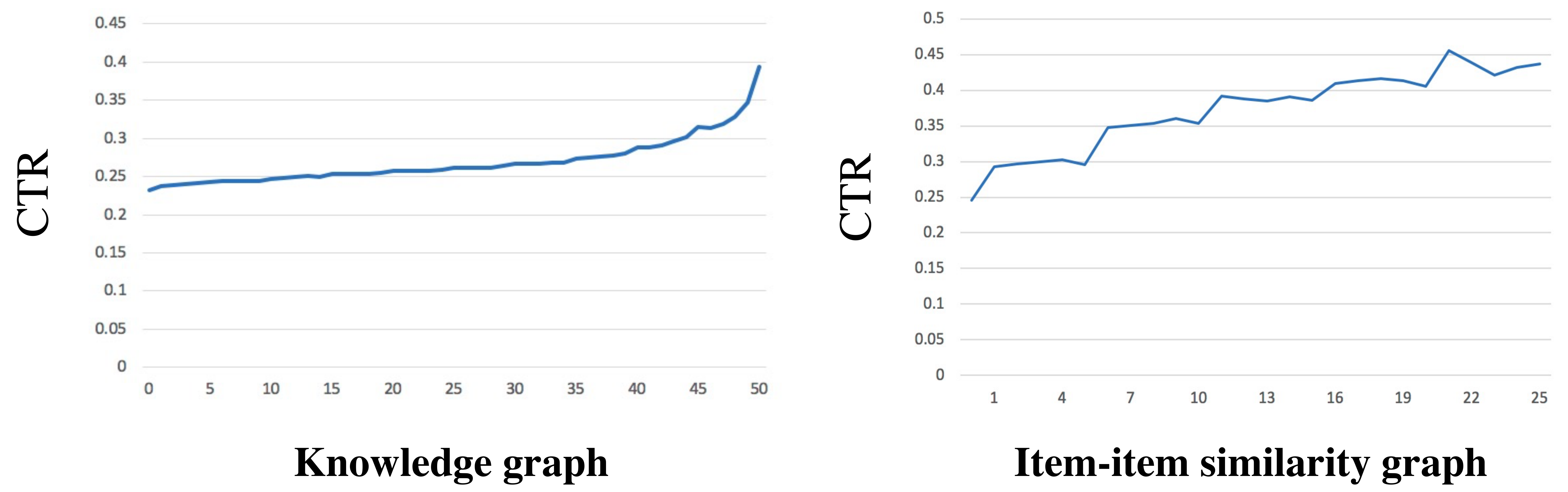}
    \end{flushleft}
    \caption{Impact of the number of relational paths.}
    \label{fig:RQ11}
\end{figure}

\begin{figure}%[htbp]
    \begin{flushleft}
    \includegraphics[scale=0.127]{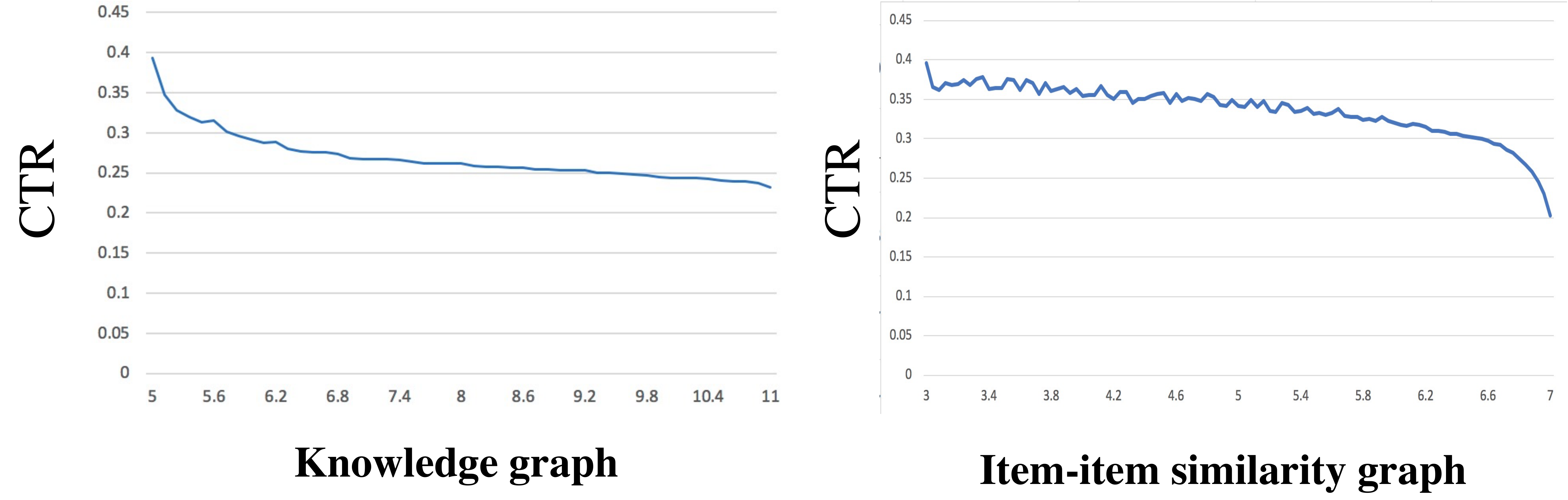}
    \end{flushleft}
    \caption{Impact of the average length of relational paths.}
    \label{fig:RQ12}
\end{figure}

In this section, we make comprehensive instance-level analyses of the effectiveness of multiplex relations (\ie the extracted relational paths) derived from different graphs on the E-commerce dataset. As shown in Figure~\ref{fig:RQ11}, CTR is calculated by averaging real label values (0 for non-click or 1 for click) of instances with the same number of paths in dataset. The number of relational paths (both on knowledge graph and item-item similarity graph) is positively correlated with CTR.
This indicates that the more relations between the target item and user behaviors, the more likely the user is to click on the target item. 
We also investigate the influence of average path length extracted from each sample on CTR. Figure~\ref{fig:RQ12} shows that, shorter relational paths from the item-item similarity graph and the knowledge graph contributes to higher CTR. It demonstrates that the shorter distance between user behaviors and the target item in a graph implies closer relation, which makes the user more likely to click on the target item.

\subsection{Online A/B Testing}
\begin{figure}
    \begin{flushleft}
    \includegraphics[scale=0.55]{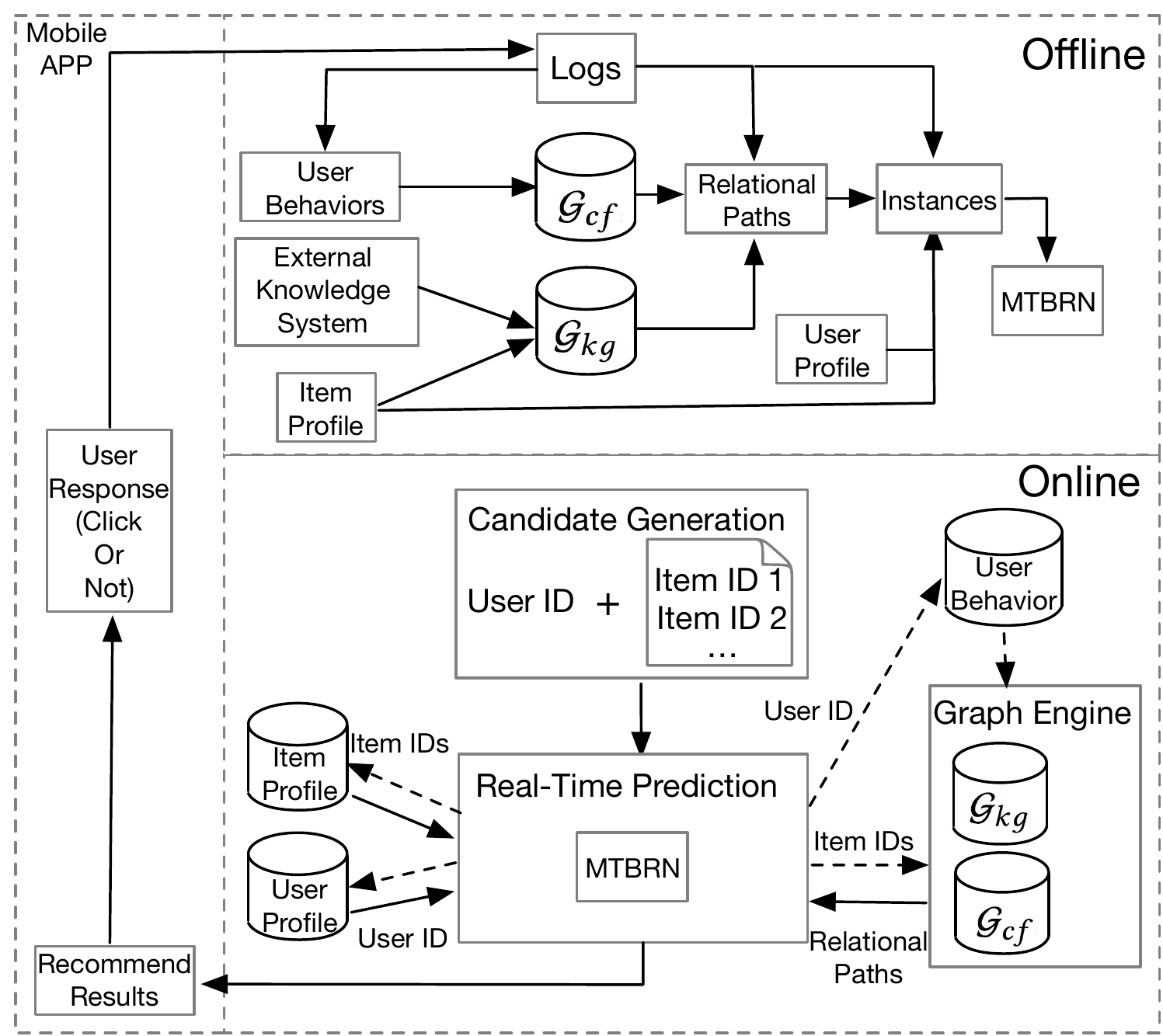}
    \end{flushleft}
    \caption{MTBRN deployment for Online CTR.}
    \label{fig:online}
\end{figure}

We have deployed MTBRN into product recommender of one popular E-commerce Mobile App for several months. The pipelines of deployment are clearly displayed in Figure~\ref{fig:online}, which consists of three parts: \textit{user response}, \textit{offline training} and \textit{online serving}.
User-item interaction logs from \textit{user response}, as well as knowledge graph and item-item similarity graph are fed into MTBRN model for training. 
When a user accesses the App, a series of candidate items are generated by MTBRN in real time. 
Subsequently, candidate items are sorted and truncated by the predicted scores, and recommended to the user.
We conducted strict online A/B testing to validate the performance of MTBRN. Our online baseline is the latest deployed DIN.
Online CTR increased by 7.9\% on average compared to DIN, at the cost of 7 milliseconds more for MTBRN online inference. 
The substantial increase in commercial revenue together with tolerable latency in serving serve as proof for the effectiveness of our proposed MTBRN.

\section{Related Work}
\subsection{User Behaviors Enhanced CTR} \label{sec:seqctr}
Click-through rate (CTR) prediction is an essential task in most industrial recommender systems. Recently, modeling user behavior sequences has attracted much attention and been widely proven effective in the CTR field. DIN~\cite{Zhou:DIN} uses the attention mechanism with embedding product to learn the adaptive representation of the user behavior sequence w.r.t. the target item. Inspired by DIN, the majority of following up works inherit this kind of paradigm.  
DIEN~\cite{zhou2019deep} and SDM~\cite{SDM2019} devote to capturing users' temporal interests and modeling their sequential relations.
DSIN~\cite{Feng:DSIN} focuses on capturing the relationships of  users' inter-session and intra-session behaviors.
Though with great improvements, the embedding product based attention mechanism fails to capture the multiplex relations between user behaviors and target item.

\subsection{Knowledge-aware Recommendation}
Many recent research studies integrate knowledge graph that integrates more side information of items into recommendation to improve the interpretability of recommendation. RippleNet~\cite{wang2018ripplenet} combines the advantages of the previously mentioned two types of methods. KPRN~\cite{Wang:KPRN} designs multiple paths between the user and item pair and uses LSTM to extract the information of each path.~\citeauthor{Wang2019KGN}~\citeN{Wang2019KGN} uses graph convolution network to automatically discover both high-order structure information and semantic information of the knowledge graph. KGAT~\cite{Wang:KGAT} leverages graph attention network to model high-order relation connectivity in the knowledge graph and user-item bipartite graph. Empirically, path-based methods make use of KG in a more natural and efficient way, that are more suitable for depicting the deep relevance between user behaviors and target item in our work for CTR prediction.

\section{Conclusion}
In this paper, we propose a new framework MTBRN to transfer the information of multiplex relations between user behaviors and target item in CTR prediction. The integration of different graphs and various connection paths ensure the superior performance of MTBRN over related work. We conducted extensive experiments on an industrial-scale dataset and a public dataset to demonstrate the effectiveness of our method. Empirical analyses show the validity of each component in the proposed framework. 

%%
%% The next two lines define the bibliography style to be used, and
%% the bibliography file.
\bibliographystyle{ACM-Reference-Format}
\bibliography{sample-base}

\end{document}